\documentclass[a4paper,11pt]{article}
\pdfoutput=1 % if your are submitting a pdflatex (i.e. if you have
\usepackage{jheppub} % for details on the use of the package, please
\usepackage[T1]{fontenc} % if needed

\usepackage{mathrsfs}
\usepackage{amsfonts}
\usepackage{color}
\usepackage{ulem}
\usepackage{graphicx}
\usepackage{amsmath}
\usepackage{cases}

\def\exd{\mathrm{d}}
\def\exD{\mathrm{D}}
\def\exi{\mathrm{i}}

\def\d{\delta}
\def\o{\omega}
\def\O{\Omega}
\def\G{\Gamma}

\def\a{\alpha}
\def\b{\beta}

\def\n{\nabla}
\def\l{\lambda}

\def\e{\epsilon}
\def\z{\zeta}
\def\zb{\bar{\zeta}}
\def\zz{\bar{z}}

\def\Lie{{\cal L}}
\def\cG{\mathcal{G}}
\def\cQ{\mathcal{Q}}

\def\cE{{\cal E}}
\def\pa{\partial}

\def\ni{\mathscr{I}}
\def\fni{\mathscr{I}^+}
\def\pni{\mathscr{I}^-}
\def\ifni{\int_{\fni} {^{(3)}}\epsilon}

\def\BMS{\mathbb{B}\text{MS}}

\def\bea{\begin{eqnarray}}
\def\eea{\end{eqnarray}}
\def\nn{\nonumber}

\title{\boldmath A note on asymptotic symmetries and soft-photon theorem}

\author{Arif Mohd}
\emailAdd{arif.mohd@sissa.it}
\affiliation{SISSA - International School for Advanced Studies, \\
via Bonomea, 265, 34136 Trieste, Italy\\ \& \\ INFN, Sezione di Trieste, Italy.}

\abstract
{ 
We use the asymptotic data at conformal null-infinity $\ni$ to formulate Weinberg's soft-photon  theorem for Abelian gauge theories with
massless charged particles. We show that the angle-dependent gauge
transformations at $\ni$ are not merely a gauge redundancy, instead they
are genuine symmetries of the
radiative phase space. In the presence of these symmetries, Poisson bracket between
gauge potentials is not well-defined. This does not pose an
obstacle for the quantization of the radiative phase space, which proceeds
by treating the conjugate electric field as the fundamental variable. 
 Denoting by $\cG_+$
and $\cG_-$ as the group of gauge transformations at $\fni$ and $\pni$
respectively, Strominger has shown that a certain diagonal
subgroup $ \cG_{diag} \subset \cG_+ \times \cG_-$ is the symmetry of the
S-matrix and Weinberg's soft-photon theorem is the corresponding
Ward identity. We give a systematic derivation of this result for Abelian
gauge theories with massless charged particles. Our derivation is a slight generalization
of the existing derivations since it is applicable even when the bulk spacetime is not exactly flat, but is only ``almost" Minkowskian.   
}

\begin{document}

%\pacs{}
%\keywords{}

\maketitle
\flushbottom
%\tableofcontents
\section{Introduction}
\label{sec:intro}
The arguments that the quantum theory of gravity should have a holographic description are rooted in black-hole thermodynamics. The latter is as true in asymptotically flat spacetimes as in the  asymptotically anti-de Sitter spacetimes. And yet, we don't have a holographic description of quantum gravity in asymptotically flat spacetimes.
In such a situation, it is reasonable to take our cue from the bottom-up approach to the established AdS/CFT dictionary \cite{Sundrum:2011ic, Kaplan:2013ads} and ask if something of the kind exists for quantum gravity in asymptotically flat spacetimes. \par
For quantum field theories in Minkowski space,  S-matrix is a holographic observable by definition. Thus there is a huge effort to calculate the S-matrix without going through the conventional Feynman calculus involving integrals in the bulk spacetime \cite{Arkani-Hamed:2013jha}. On the other hand, recently there has been a flurry of activity following the realization by Strominger \cite{Strominger:2013lka, Strominger:2013jfa} that Weinberg's soft theorems of gauge
theories and gravity can be seen as the tree-level Ward identity for a symmetry
of the S-matrix \cite{He:2014cra, He:2014laa, Cachazo:2014fwa, Lysov:2014csa, Kapec:2014opa}. For gravity, the symmetry in question is a certain diagonal subgroup of the product group of supertranslations on the future and past null infinity, $\fni$ and $\pni$, which comprise the null boundary of asymptotically flat spacetimes. For gauge theories, the symmetry in question is a certain
diagonal subgroup of the product group of gauge transformations on $\fni$ and $\pni$. Strominger has further
conjectured on the form of the subleading terms in the soft-theorem, and it was verified in Ref.~\cite{Cachazo:2014fwa}. The analogous terms in the case of Yang-Mills theories were found in Refs.~\cite{Casali:2014xpa, Larkoski:2014hta}. The precursor of these developments was the proposal by Barnich and Troessaert~\cite{Barnich:2009se, Barnich:2010eb,
Barnich:2011ct, Barnich:2011mi} (see also Refs.~\cite{Banks:2003vp, Barnich:2013axa, Barnich:2013sxa}) that the group of asymptotic symmetries of asymptotically flat spacetimes should include the infinite dimensional Virasoro subgroup. Therefore, a general expectation is that the techniques familiar from two-dimensional conformal field theories could be imported in developing a holographic description of quantum gravity in asymptotically flat space. \par
 With holography in flat space as the main motivation, in this paper we revisit the Weinberg's soft-photon theorem in Abelian gauge theories with massless charged matter studied in Refs.~\cite{Strominger:2013lka, He:2014cra} and  formulate it in terms of the
quantities defined intrinsically on the null boundary of a conformally compactified asymptotically flat spacetime. Our goal in this paper is to systematically study the asymptotic symmetries of the radiative phase space; to quantize the phase space and; following  Refs.~\cite{Strominger:2013lka, He:2014cra}, to derive the soft-photon theorem as a Ward identity related to large gauge-symmetry of the radiative  phase space. In line with the holographic motivation, our work will be exclusively on the null boundary  $\ni$ of spacetime and the bulk is discarded after the construction of the radiative phase space. For studies related to the construction of a theory living on $\ni$ that calculates the tree level amplitudes in supergravity see Refs.~\cite{Adamo:2014yya, Geyer:2014lca}. \par
 Our derivation of the soft-theorem as a Ward identity is a slight generalization of the derivations in Refs.~\cite{Strominger:2013lka, He:2014cra} in the sense that ours is applicable even when the bulk spacetime is not exactly flat but is ``close" to Minkowski in some sense. We, however, do not consider the backreaction, thus the bulk is still non-dynamical. During this study we will uncover an interesting subtlety related to the Poisson brackets on the radiative phase that was also noted in Ref.~\cite{Campiglia:2014yka}. This would lead us to identify soft-photons as ``edge states" on $\ni$.  \par
This paper is organized as follows-- in Sec.~\ref{sec:scri} we review the definition 
of asymptotic flatness and the geometry of  null infinity $\ni$; 
in Sec.~\ref{sec:softamp} we review
Weinberg's soft-photon theorem and express it in the coordinates intrinsic to $\ni$; 
in Sec.~\ref{sec:em} we introduce the radiative phase space of electromagnetism and 
quantize it; in Sec.~\ref{sec:phtoS} we review Strominger's proposal for new symmetries of 
the scattering problem in massless quantum electrodynamics; in Sec.~\ref{sec:wardsoft}
 we impose the invariance of the S-matrix under large
gauge symmetry to derive the Weinberg's soft-photon theorem. We conclude with
a summary and outlook in Sec.~\ref{sec:summary}. A 
brief review of the symplectic formulation of field
theory, and the example of free Maxwell field, is provided in the appendix.\par
We will be working in four-dimensional bulk with the metric-signature $(-+++)$. The variation symbol $\d$ will represent
the exterior derivative on phase space. We will leave the ``$\wedge$" symbol implicit in the differential
forms on  phase space.
\section{Geometry of null infinity} 
\label{sec:scri}
  In this section
we review the geometry of null infinity $\ni$. Since $\ni$ arises as the conformal  boundary of asymptotically flat spacetimes \cite{Penrose:1962ij}, we begin with the definition of asymptotic flatness. We will be following the recent review in Ref.~\cite{ashtekar2014geometry}. \par
 A spacetime $(\hat{M}, \hat{g}_{ab})$ is said to be asymptotically flat at null infinity if
there exists a manifold $M$ with boundary, $\partial M := \ni$, equipped with a
smooth metric of signature $(-+++)$ such that the interior $M-\ni$ is
diffeomorphic to $\hat{M}$, and the following conditions are satisfied:
\begin{enumerate} \item there exists a smooth function $\O$ on $M$ such that
$g_{ab} = \O^2 \hat{g}_{ab}$ on $\hat{M}$, with $\O=0$ and $\partial_a \O \neq
0$ on $\ni$, \item $\ni$ is topologically $\mathbb{S}^2 \times \mathbb{R}$, the
vector field $n^a := g^{ab} \partial_b \O$ on $\ni$ is complete and the space of
its orbits is diffeomorphic to $\mathbb{S}^2$, \item $\hat{g}_{ab}$ satisfies the Einstein equation in
the intersection of $\hat{M}$
with a neighborhood of $\ni$ in M and $\O^{-2} \hat{T}_{ab}$ extends smoothly to $\ni$.
\end{enumerate}
The intrinsic geometry of $\ni$ is described by a degenerate metric $q_{ab}$ and the null generator $n^a$.
There is a conformal freedom in the choice of the rescaled metric $g_{ab}$,
 \bea
 \O \rightarrow \o \O \,\,\, , \,\,\, \Lie_n \o = 0,
\eea
which leads to the following freedom in the intrinsic metric and the null generator of $\ni$,
\bea
  \{q_{ab},n^a\} \rightarrow \{\o^2 q_{ab}, \o^{-1} n^a\}.
\eea
Therefore, the universal structure at $\ni$ consists of the equivalence class
  $ \{q_{ab},n^a\} \sim \{\o^2 q_{ab}, \o^{-1} n^a\}$. The vector fields
on $\ni$ that respect this universal structure are the infinitesimal generators
of the asymptotic symmetry group of asymptotically flat spacetimes. This group
is called the $\BMS$ \footnote{The bold  $\mathbb{B}$ with respect to 'MS' is to emphasize that $\mathbb{B}$ stands for the names of two authors.}
 group after the founders Bondi, van der Burg, Metzner \cite{Bondi:1962px} and  Sachs \cite{Sachs:1962zza}. The vector fields $\xi^a$ that generate the $\BMS$ group satisfy, 
\bea
\Lie_\xi q_{ab} = 2 \b q_{ab} \,\,\,\, \text{and} \,\,\, \Lie_\xi n^a =-\b n^a, 
\eea
where $\b$ is a scalar on $\ni$ such that $\Lie_n \b = 0$. Those $\xi^a$ which are of the form $\xi^a=f\, n^a$, where $f$ is a function such that $\Lie_n f = 0$, generate the normal subgroup of the $\BMS$ group and are called supertranslations (ST). These vector fields generating ST form a Lie ideal, in the sense that their commutator with any $\BMS$ vector field is again a ST. The quotient group $\BMS/ST$ is generated by the conformal isometries of $\mathbb{S}^2$ and is isomorphic to the Lorentz group. 
As originally formulated in Refs.~\cite{Bondi:1962px, Sachs:1962zza} the conformal isometries of $\mathbb{S}^2$ were assumed to be complete vector fields.
Recently, Barnich and Troaessart \cite{Barnich:2009se, Barnich:2010eb,
Barnich:2011ct, Barnich:2011mi} have proposed that the singular vector fields should also be allowed and in that case the quotient $\BMS/ST$ is no longer
the finite dimensional Lorentz group, but it is the infinite dimensional Virasoro group. This observation implies that the techniques from two-dimensional conformal field theories could play an important role in the physics of four dimensional asymptotically flat spacetimes.\par
We will work in the Bondi gauge, so that $q_{ab}$ is the metric of a unit $\mathbb{S}^2$, $\exd s^2 = \exd \theta^2 + r^2 \sin^2 \theta \,\exd \phi^2$. By using stereographic projection from the North-pole of $\mathbb{S}^2$ to the equatorial plane, we can label the coordinates on the unit $\mathbb{S}^2$ by a complex number $\z = \cot(\theta/2) \mathrm{e}^{i\phi}$. Thus we represent a  point on the unit $\mathbb{S}^2$ by coordinates $\{\z, \zb\}$, and we view $\zb$ not as a complex conjugate of $\z$ but as an independent coordinate. The cartesian components of a point $\{x=\sin  \theta \cos \phi,  y=\sin \theta \sin \phi, z=\cos \theta\}$  on the unit $\mathbb{S}^2$ are then given by
\bea
\label{eq:cartcomp}
x = \frac{\z + \zb}{\z \zb + 1}, \,\,\,\, 
y = \frac{1}{i}\frac{\z - \zb}{\z \zb + 1}, \,\,\,\,
z =\frac{\z \zb - 1}{\z \zb + 1}. 
\eea

The metric on unit $\mathbb{S}^2$ in terms of the $\{\z , \zb\}$ is given by
\bea
\label{eq:stereometric}
\exd s^2 = \frac{4}{(1+\z \zb)^2} \exd \z \exd \zb.
\eea
\section{Amplitudes with soft-photon insertion}
\label{sec:softamp}
In this section we first review Weinberg's soft-photon theorem in Sec.~\ref{subsec:Wsoft}.  Our treatment follows that of Ref.~\cite{Weinberg:1995mt}.\footnote {For the derivation of soft-factors in gauge theories  using the soft-collinear effective field theory please see Ref.~\cite{Larkoski:2014bxa}.} Then in Sec.~\ref{subsec:sfonni} we rephrase the soft-factor that appears in the soft-photon theorem in terms of the quantities defined intrinsically on $\ni$. 
\subsection{Weinberg's soft-photon theorem}
\label{subsec:Wsoft}
Weinberg's soft-photon theorem \cite{Weinberg:1965nx, Weinberg:1995mt} is a universal formula that gives the amplitude for emission
of arbitrary number of very low-energy photons in a process $\alpha \rightarrow \beta$ involving any
number of higher-energy charged particles of any kind in the initial state $\alpha$ and the final state $\beta$. In particular, the amplitude $\mathscr{M}^{\mu}_{\alpha \rightarrow \beta}$ for emitting a single soft-photon with four-momentum $q$ and polarization index $\mu$ in the process $\alpha \rightarrow \beta$ is given in the soft limit (i.e., the limit in which the energy of the photon $\o \rightarrow 0$) as 
\bea
\label{eq:Wsoftfund}
\lim_{\o \rightarrow 0} \mathscr{M}^{\mu}_{\alpha \rightarrow \beta}(q)  = \mathscr{M}_{\alpha \rightarrow \beta} \left[ \sum_{n \in \beta} e_n\frac{p^\mu_n}{q \cdot p_n}
- \sum_{n \in \alpha} e_n \frac{p^\mu_n}{q \cdot p_n} \right],
\eea
where $p_n$ and $e_n$ are the  four-momentum and the charge of the $n^{th}$ particle in the $in$ ($out$) state $\a$ ($\b$), $\o$ and $q$  are the energy and four-momentum of the emitted soft-photon, and the dot ``$\cdot$" stands for the index- contraction. The quantity in parenthesis on the right-hand-side is the tree-level soft-factor at the leading order.  \par
One interesting implication of Eq.~\eqref{eq:Wsoftfund} is obtained by contracting it with the photon four-momentum $q$ and demanding that the result should vanish in order to  preserve the Lorentz-invariance (since the polarization ``vector" is not really a four-vector and we could choose it upto the addition of the photon's four-momentum). This gives $\sum_{n \in \beta} e_n = \sum_{n \in \alpha} e_n$, i.e.,  the total charge of the system is conserved. \par
We will be interested in the amplitude for emitting a soft-photon for a particular polarization, say $\e^{(+)}_{\mu}$, which is obtained by contracting Eq.~\eqref{eq:Wsoftfund} with   $\e^{(+)}_{\mu}$, 
\bea
\label{eq:Wsoft}
\lim_{\o \rightarrow 0} \mathscr{M}^{(+)}_{\alpha \rightarrow \beta}(q)  = \mathscr{M}_{\alpha \rightarrow \beta} \left[\sum_{n \in \beta}  e_n\frac{\e^{(+)} \cdot p_n}{q \cdot p_n}
- \sum_{n \in \alpha} e_n \frac{\e^{(+)} \cdot p_n}{q \cdot p_n} \right].
\eea
The amplitudes appearing on the two sides of Eq.~\eqref{eq:Wsoft} are asymptotic observables, i.e., their natural habitat is $\ni$. We therefore turn to recast the soft-factor in terms of the quantities intrinsically defined on $\ni$ so that the soft-theorem could be expressed intrinsically on $\ni$.
\subsection{Soft-factor on $\ni$}
\label{subsec:sfonni}
In this sub-section, we rewrite the soft-factor, which appears on the right-hand-side of Eq.~\eqref{eq:Wsoft}, in the Bondi gauge  and in terms of the stereographic coordinates on the $\mathbb{S}^2$ cross-sections of $\ni$ introduced in Sec.~\ref{sec:scri}. \par Denote the four-momentum of a particle as $p^\mu = \{E, \vec{p}\}$. For a massless particle, $E = |\vec{p}|$. Hence, 
$p^\mu = E\{1,\hat{p}\}$ for a massless particle. But $\hat{p}$ points in the direction of motion of the particle. If we assume that the scattering happens at the center of the two-sphere at infinity then the direction of motion of all the particles involved is radial. In this approximation, $\hat{p}$ can be characterized in terms of the point on the unit $\mathbb{S}^2$ cross-section of  $\ni$ where the particle finally hits. The latter is  coordinatized by $\{\z, \zb\}$. Thus the four-momentum of the massless particle can be encoded using the stereographic coordinates as $\{E, \z , \zb \}$. \par 
The dot product between two null-momenta $p_1 = \{E_1, \z_1 , \zb_1 \}$ and $p_2 = \{E_2, \z_2 , \zb_2 \}$ can be calculated using Eq.~\eqref{eq:cartcomp} to be,
\bea
\label{eq:dotnull}
p_1 \cdot p_2 = -2 E_1 E_2 \frac{(\z_1 - \z_2)(\zb_1 - \zb_2)}{(1+\z_1 \zb_1)(1+\z_2 \zb_2)}. 
\eea
For a photon we also need polarization vectors corresponding to a given four-momentum.  Following Ref.~\cite{He:2014cra}, we take the two independent transverse polarization vectors corresponding to the photon $q=\{\o,\z, \zb \}$ to be
\begin{subequations}
\label{eq:pol}
\bea
\e^{(+)\mu}&=&\frac{1}{\sqrt{2}}\{\zb, 1, -i, \zb \}, \label{eq:polplus} \\
\e^{(-)\mu}&=&\frac{1}{\sqrt{2}}\{\z, 1, i, \z \}.  \label{eq:polminus}
\eea
\end{subequations}
Projecting the polarizations \eqref{eq:pol} on the conformal $\mathbb{S}^2$ at $\ni$ using $\e_a = \{ \dfrac{\pa x^\mu}{\pa \z} \e_\mu, \dfrac{\pa x^\mu}{\pa \zb} \e_\mu \} $ we get the induced polarizations on $\mathbb{S}^2$,
\begin{subequations}
\label{eq:polscri}
\bea
\e_a^{(+)} &=& \{0,\frac{\sqrt{2}}{(1+\z \zb)} \}, \label{eq:polscriplus} \\
\e_a^{(-)} &=& \{\frac{\sqrt{2}}{(1+\z \zb)}, 0 \}. \label{eq:polscriminus}
\eea
\end{subequations}
%Note that, by definition, $\e^+_\z = \dfrac{\pa x^\mu}{\pa \z} \e^+_\mu$ and $(\e^+_\mu)^{\ast} = \e^-_\mu$.  Therefore, $(\e^+_\z)^{\ast} = \dfrac{\pa x^\mu}{\pa \zb} \e^-_\mu = \e^-_{\zb} $.
The dot-product between the polarization vectors and null momenta $p=\{z, \zz \}$ can be calculated by writing the momenta explicitly in terms of its components using Eq.~\eqref{eq:cartcomp}. 
Then, using Eqs.~\eqref{eq:dotnull} and \eqref{eq:polplus}, for the soft-factor in Eq.~\eqref{eq:Wsoft} we obtain,
\bea
\left[ \sum_{n \in \b} e_n\frac{\e^{+} \cdot p_n}{q \cdot p}
- \sum_{n \in \a} e_n \frac{\e^{+} \cdot p_n}{q \cdot p} \right] = \frac{(1+\z \zb)}{\sqrt{2}\o}  \left[ \sum_{n \in \b}e_n\frac{1}{\z - z_n} - \sum_{n \in \a} e_n\frac{1}{\z - z_n} \right],
\eea
where the $n^{th}$ charged particle with four-momentum $p^\mu_n$ has coordinates $\{E_n, z_n,\zz_n\}$. If we label the asymptotic momenta in the ingoing (outgoing)  states $\alpha$ ($\beta$)  by their corresponding stereographic coordinates on $\mathbb{S}^2$ at $\fni$ ($\pni$)  and their asymptotic energy as $\{E, z, \zz \}$, then the soft-theorem gets expressed in terms of quantities defined intrinsically on $\ni$ as
\bea
\label{eq:softni}
\lim_{\o \rightarrow 0} \mathscr{M}^{(+)}_{\alpha \rightarrow \beta}(q)  = \mathscr{M}_{\alpha \rightarrow \beta} \frac{(1+\z \zb)}{\sqrt{2}\o}  \left[ \sum_{n \in \b}e_n\frac{1}{\z - z_n} - \sum_{n \in \a} e_n\frac{1}{\z - z_n} \right].
\eea
Thus the natural habitat of the soft-theorem is also $\ni$. Eq.~\eqref{eq:softni} is the form of Weinberg's soft-photon theorem that we will derive in Sec.~\ref{sec:wardsoft}.
\section{QED with massless matter}
\label{sec:em} 
In this section we first review the construction of the phase space of radiative modes in classical electrodynamics with massless matter, then we quantize it. Our classical consierations follow Ref.~\cite{Ashtekar:1987tt} (see also, Ref.~\cite{Ashtekar:1981bq}). Quantization of radiative modes on a null surface was also studied in Ref.~\cite{Frolov:1977bp}. 
\subsection{Classical phase space}
\label{subsec:phspace}
Let's denote the pullback of the
electric field $E_a = F_{ab}n^b$ and the rescaled matter current $\O^2 J^a$ to
$\fni$ as $\cE_a$ and $j^a $, respectively. Let $A_a$ be a vector potential for
the pullback to $\fni$ of the field tensor $F_{ab}$. These quantities determine the electromagnetic field $F_{ab}$
everywhere in $({M},{g}_{ab})$, the conformal completion of
$(\hat{M},\hat{g}_{ab})$ \cite{Ashtekar:1987tt}. Since, $\cE_a n^a = 0$, we have only two independent
components in $\cE_a$. We recognize these components as the two radiative modes of the Maxwell field.
The symplectic structure is given by 
\bea 
\label{eq:maxrsymp} 
\O = \frac{1}{4\pi} \ifni \,q^{ab}\,\d \cE_a \, \d A_b. 
\eea 
In the gauge $A_a n^a =
0$, using that $\exD_a n^b = 0$, we also have that $\cE_a=\Lie_n A_a$. Note that unlike  Ref.~\cite{Ashtekar:1987tt} we do not require that $A_a \to 0$ at the future and past boundary of $\ni$ ($\fni_\pm$). We still
have the residual gauge freedom $A_a \rightarrow A_a + \exD_a \l$, where $\Lie_n
\l = 0$, i.e., $\l$ is a function on the space of generators of $\fni$. This gauge freedom is a genuine symmetry and not merely a redundancy in our description. The vector field $X_\l = \ifni \,\, \exD_a
\l \dfrac{\d}{\d A_a}$, which corresponds to the residual gauge freedom, is not a degenerate direction of $\O$. To see this, let us
calculate the inner product 
\bea \exi_{X_\l} \O &=& \frac{1}{4 \pi} \ifni \,\,
q^{ab} \, \left[\Lie_n (\exD_a \l)\, \d A_b - \d \cE_a \, \exD_b \l \right] \nn \\
 &=& \frac{-1}{4 \pi} \ifni \,\, q^{ab} \,
\Lie_n(\d A_a) \, \exD_b \l \nn \\ &=& \frac{-1}{4 \pi} \ifni \, \, q^{ab}
\,\Lie_n(\d A_a \, \exD_b \l \,) \nn \\ &=& \frac{-1}{4 \pi} (\int_{\fni_+} -
\int_{\fni_-}) ^{(2)}\epsilon \,q^{ab} \, \d A_a \, \exD_b \l, 
\eea 
where in the
first step we have used that $\cE_a = \Lie_n A_a $; in the second that $\Lie_n
\l = 0, \Lie_n q^{ab} = 0$ and $\exD_a n^b = 0$; and in the third we have used
the Stokes theorem to get the integrals at the future and the past boundaries of
$\fni$ that, following Strominger in Ref.~\cite{Strominger:2013lka}, we will denote by ${\fni_+}$ and ${\fni_-}$, respectively. Finally, the charge $Q_\l$ that generates the residual gauge transformation with gauge parameter $\l$ can be calculated by integrating 
$\exi_{X_\l} \O = \d Q_\l$. This gives,
 \bea
 Q_\l = \frac{-1}{4 \pi}
(\int_{\fni_+} - \int_{\fni_-}) ^{(2)}\epsilon \,\, q^{ab} \,\, A_a \, \exD_b
\l.
 \eea 
The gauge transformations which are non-vanishing at infinity, in particular at $\fni$ are called the large gauge transformations \cite{Strominger:2013lka}. We thus see that the large gauge transformations are the symmetries of the phase space. Henceforth, we will call it the large gauge-symmetry. \par
Following Ref.~\cite{He:2014cra}, we now assume that the field
strength vanishes at boundaries $\fni_+$ and $\fni_-$, i.e., the
vector potential is pure gauge there. Let 
\bea A_a &=& \exD_a \phi_+ \,\,\,\,\,
\text{at} \,\,\,\,\, \fni_+ \\ A_a &=& \exD_a \phi_- \,\,\,\,\, \text{at}
\,\,\,\,\, \fni_-. 
\eea
 We can map $\fni_-$ to $\fni_+$ by following the integral curves of the null-generator $n^a$. Thus we can  think of $\phi_- $ as a function on
$\mathbb{S}^2$ at $\fni_+$ and hence write the charge as
\bea Q_\l &=& \frac{-1}{4 \pi} \int_{\fni_+} {^{(2)}}\epsilon \, \, q^{ab} \,\,
\exD_a (\phi_+ - \phi_-)\, \exD_b \l, \label{eq:chargesd} \\ &=& \frac{1}{4 \pi} \int_{\fni_+}
{^{(2)}}\epsilon \, \, \l \,\, q^{ab} \,\, \exD_a \exD_b (\phi_+ - \phi_-), \label{eq:chargedd}\eea
where in the second line we integrated by parts. The charge can also be written as an integral over the whole of
$\fni$ as
 \bea 
\label{eq:chargeE}
Q_\l = \frac{-1}{4 \pi} \ifni \,\, q^{ab} \, \cE_a \, \exD_b \l.
\eea 
The algebra of generators of the large gauge-symmetry is abelian, essentially because $\cE_a$ is gauge invariant,
\bea
\label{eq:gaugechargealgebra}
\left\{Q_{\l_1}, Q_{\l_2} \right\} = 0.
\eea 
In the Bondi gauge, using integration by parts in Eq.~\eqref{eq:chargedd}
we can also write the charge as 
\bea
\label{eq:chargel}
Q_\l = \frac{-1}{2 \pi} \int_{\fni} \exd u \, \exd^2\z \sqrt{q}  \,\, q^{\z \zb} \, \cE_{\z} \, \exD_{\zb} \l.
\eea
Now plugging in the explicit form of metric, choosing the gauge parameter $\l$ as $\l = \frac{1}{\z - z}$, and using the identity $\pa_{\bar{z}} (1/(z-\z)) = 2 \pi \d (z-\z,\bar{z}-\zb)$ in Eq.~\eqref{eq:chargel} we get
\bea
\label{eq:quantizableQ}
Q_{\l = \frac{1}{\z - z}} = -\int_{\fni} \exd u \,\,  \cE_{z}(u, z, \bar{z}),
\eea
which is the form that we will eventually use for quantizing this operator in Sec.~\ref{subsec:quantizeE}.
It is clear from Eq.~\eqref{eq:quantizableQ}  that the generator of large gauge-symmetry for the gauge
parameter $\l=\frac{1}{\z - z}$ is the zero mode of the $z$-component of $\cE$ \cite{He:2014cra}. Similarly, the generator 
of large gauge-symmetry for the gauge parameter $\l=\frac{1}{\zb - \bar{z}}$ is the zero mode of the $\bar{z}$
component of $\cE$. \par
Before moving on to quantize the phase space, for the sake of completeness and for potential future use, 
we study the action of the $\BMS$ symmetry on the radiative phase space. 
\subsection{Action of $\BMS$ symmetry} 
\label{subsec:bmsaction}
Let $\xi^a$ be a $\BMS$ vector
field. Hence, \bea \Lie_\xi q_{ab} = 2 \b q_{ab} \,\,\,\, \text{and} \,\,\,
\Lie_\xi n^a =-\b n^a, \eea where $\b$ is a scalar such that $\Lie_n \b = 0$.
The vector field $\xi$ induces a vector field $X_\xi$ on the phase space
that is given by \bea X_\xi = \ifni \,\,\Lie_\xi A_a \frac{\d }{\d A_a}. \eea To calculate the
charge generating the motion along $X_\xi$ we calculate 
\bea 
\d Q_\xi &=& \exi_{X_\xi} \O \\ 
&=& \frac{1}{4 \pi} \ifni \,\, q^{ab} \,\left[ \Lie_n (\Lie_\xi) A_a \d A_b - \d \cE_a \, 
\Lie_\xi A_b \right],
\eea
 which can be integrated using that $\cE_a \to 0$ on both the boundaries $\fni_{\pm}$ of $\fni$
, and that $\Lie_\xi \, ^{(3)}\e = 3 \b \,^{(3)}\e$, to get the charge as noted in Ref.~\cite{Ashtekar:1981bq}, 
\bea
 Q_\xi = \frac{-1}{4 \pi} \ifni \,\, q^{ab} \,
\Lie_n A_a \, \Lie_\xi A_b.
 \eea 
 In particular, for the vector fields generating
supertranslations, i.e., $\xi^a = f n^a$ where $f$ is some function such that
$\Lie_n f = 0$, the corresponding charge is \bea Q_\xi &=& \frac{-1}{4 \pi}
\ifni \,f \, q^{ab} \, \Lie_n A_a \, \Lie_n A_b \\ &=& \frac{-1}{4 \pi} \ifni
\,f\, q^{ab} \, \cE_a \, \cE_b. \eea 
%{\color{red} Check the bracket between $Q_\xi$ and $Q_\l$.} 
We have also checked that the $\BMS$ charge algebra is closed, i.e., for two $\BMS$ vector fields $\xi_1$ and $\xi_2$  we have,
\bea
\left\{Q_{\xi_1}, Q_{\xi_2} \right\} = Q_{[\xi_1,\xi_2]},
\eea
where $[\cdot, \cdot]$ is the Lie-bracket.
The bracket between the generators of large-gauge symmetry and  $\BMS$ charges can also
be calculated and is given by,
\bea
\left\{Q_{\l}, Q_{\xi} \right\} = Q_{\Lie_\xi \l},
\eea
where $\l$ is a large gauge parameter. In particular, large gauge-symmetry generator commutes with the generator of supertranslations.
Furthermore, large-gauge symmetries and $\BMS$ transformations together form a Lie algebra, with
large-gauge transformations forming another abelian ideal, one being the supertranslations.
\subsection{Quantization: $A$-representation}
\label{subsec:quantizeA}
  It is tempting to read-off the Poisson brackets for the vector potential from the symplectic structure in \eqref{eq:maxrsymp} as
\bea
\{A_a(u,\z,\zb), A_b(u',\z',\zb') \} =\frac{\hbar}{4 \pi i} q_{ab}\,\, \Delta(u-u')
 \d(\z-\z';\zb-\zb'),
\eea
where $\Delta$ is the step-function. But we now demonstrate that the Poissson bracket between $A$'s is in fact not well-defined. This was also noted in Ref.~\cite{Campiglia:2014yka} in the gravitational context in the study of the sub-leading soft-graviton theorem.
For the ease of presentation, we consider the simple toy model given in Ref.~\cite{Campiglia:2014yka} by ignoring the angular variables in the 
symplectic structure \eqref{eq:maxrsymp}.
\subsubsection*{A toy model}
 Consider a field theory with the following symplectic structure,
 \bea
 \O = \int \exd u \,\d \phi \, \d \dot{\phi}, %\,\,\,\, \d \phi \neq 0  \,\, \text{at} \,\, \infty. 
 \eea
such that the field $\phi$ and its variation $\d \phi$ do not vanish as $u \rightarrow \pm \infty$, while $\dot{\phi}(u)$ does vanish as $u \rightarrow \pm \infty$. Here overdot stands for the derivative with respect to $u$.
 It should be noted that, modulo the angular dependence, this is essentially the same symplectic structure as we have on the radiative phase space in Sec.~\ref{subsec:phspace} with $\phi$ for $A_a$ and $\dot\phi$ for $\cE_a$.
 In order to find the Poisson bracket $\{\phi(u'),\phi(u'')\}$ we first need to calculate the Poisson bracket between two functionals $\{F[\phi],G[\phi]\}$, where
 \bea
F[\phi] & = &\int \exd u\,\, f(u) \phi(u),  \nonumber\\ 
G[\phi] &= &\int \exd u \,\, g(u) \phi(u), \nonumber 
\eea
and then put $f(u) = \delta (u-u')$ and $g(u) = \delta (u-u'')$. We now show that if we do this then it would lead to a contradiction. \par
  Let $X_F$ be the vector field generated by the function $F[\phi]$ on phase space, i.e.,    
\bea
X_F=\int \exd u\,\,  X(u) \frac{\d}{\d \phi(u)}. \nonumber 
\eea
 By definition, we have $\d F = \exi_{X_F} \O$. This gives
\bea %\d F &=& \exi_{X_F} \O \nonumber \\
  \int \exd u\,\,  f(u) \d \phi(u) &=& X(u)\,  \d \phi(u) |_{-\infty}^{+\infty} - 2 \int \exd u\,\,  \dot{X}(u)\, \d \phi (u). \nonumber 
\eea
Since $\d \phi(u)$ is arbitrary, with $\d \phi (\pm \infty) \neq 0$, we must have that
\bea
  %&\implies& X \,\,\,\text{ is Hamiltonian only if} \,\,\, X(\pm \infty) = 0. \nonumber \\
  X(\pm \infty) = 0 \,\,\,\, &\text{and}& \,\,\,\,f(u) = -2\dot{X}(u) , \nonumber \\
  &\implies& \int_{-\infty}^{+\infty} \exd u \, f(u) = 0. \,\,\,\nonumber
\eea
But this is not compatible with  $f(u)$ being a delta function $\d (u - u')$, which
we need to calculate the bracket $\{\phi(u'),\phi(u'')\}$. We conclude that this Poisson bracket  is not well-defined. It can also be checked that the Poisson bracket between $\dot{\phi}(u)$ is actually well-defined since $\dot{\phi}(u) \to 0$ as $u\to \pm \infty$. In the same vein, Poisson bracket between vector potentials in our massless abelian theory is also not well-defined, but the Poisson bracket between electric fields is well-defined. Thus it is not surprising that a 
naive extraction of the Poisson bracket between vector potentials from the symplectic structure yields ambiguous results, such as the ``factor of $1/2$"
problem encountered in Refs.~\cite{He:2014cra, He:2014laa}. One way to cure this problem is to enlarge the phase space by introducing new degrees of freedom so that the contradiction noted above is avoided. These degrees of freedom are also called the ``edge states" (see, for e.g.,  Refs.~\cite{Balachandran:1994vi, Balachandran:1992rb}).
This is the procedure followed in Refs.~\cite{He:2014cra, He:2014laa}. Let us now see how it works in our toy model. \par 
We introduce a new degree of freedom $\phi_{-} $, which is really the boundary value of the field $\phi$ at $u \rightarrow -\infty$ (hence the name -- edge state).  Let us take as its symplectic partner $\phi_{+}$, which is the boundary value of $\phi$ at $u\rightarrow +\infty$.  Symplectic structure on the enlarged phase space is now taken to be
\bea
 \O = \int \exd u \,\d \phi \, \d \dot{\phi} + \d \phi_{-} \d \phi_{+}. \nn
\eea
In this enlarged phase space,  $\d F  = \exi_{X_F} \O$ gives
\bea
  \int \exd u\,\,  f(u) \d \phi(u) = X(u)\,  \d \phi(u) |_{-\infty}^{+\infty} - 2 \int \exd u\,\,  \dot{X}(u)\, \d \phi (u) + X(-\infty) \d \phi_+ - X(\infty) \d \phi_+, \nonumber \\
= - 2 \int \exd u\,\,  \dot{X}(u)\, \d \phi (u) + \left( X(\infty)+X(-\infty) \right)(\d \phi_+ - \d \phi_-). \nn 
\eea
 If we impose anti-periodic boundary conditions on $X(u)$ then we are left with $f(u) = -2  \dot{X}(u) $ as before, which after integration gives $\int_{-\infty}^{+\infty} \exd u \, f(u) = -2 (X(\infty)-X(-\infty))$. Since $X(\pm \infty)$ is not constrained to vanish anymore, we can choose its value consistent with the choice of delta function for $f(u)$. On the enlarged phase space, the Poisson bracket $\{\phi(u'),\phi(u'')\}$ can be read-off from the symplectic structure without causing any inconsistency anywhere. This is the procedure followed in Refs.~\cite{He:2014cra, He:2014laa}, where the new degree of freedom (the edge state) $\phi_{-}$ is identified as the Goldstone mode. \par
We are going to follow another route. In the original phase space, although the Poisson brackets between $A$'s is not well defined, the Poisson brackets between the electric field $\cE$'s is well-defined.  Therefore, without further ado we move on to quantize the phase space in the $\cE$-representation.
\subsection{Quantization: $\cE$-representation} 
\label{subsec:quantizeE}
 Following Ref.~\cite{Ashtekar:1987tt} (see also Ref.~\cite{Frolov:1977bp}), commutation relations of $\cE$'s can be read-off from the
symplectic structure in Eq.~\eqref{eq:maxrsymp} as 
\bea 
\label{eq:Eccr}
\left[ \cE_a(u, \z, \zb), \cE_b(u', \z', \zb') \right]
= \frac{1}{4 \pi \exi} q_{ab}\, \d(\z-\z',\zb-\zb')\, \Delta(u-u'),
 \eea
where $\Delta$ is the step function.
We expand the electric field in the positive- and negative-frequency Fourier
modes,
 \bea 
\label{eq:Efourier}
\cE_a(u, \z, \zb) = \int_0 ^\infty \exd
\o\,\o \left[\bar{\e}^\a_a \, a_\a(\o, \z, \zb) \, e^{-\exi \o u} + \e^{\a}_a  \,
b_\a(\o, \z, \zb) \, e^{\exi \o u}\right],
\eea 
where we take the polarizations to satisfy $\bar{\e}^\a_a \, \e^\b_b \, \d_{\a \b} = q_{ab}$, and $\bar{\e}^\a_b$ is defined such that $\bar{\e}^{\pm}_a=\e^{\mp}_a$. The reason for choosing the multiplicative factor of $\o$ with operators $a_\a, b_\a$ is that we want to associate these operators with the creation/annihilation of photons, which are particles corresponding to the vector-potential $A_a$. Since $\cE_a = \Lie_n A_a$, the expansion of the operator $A_a$ would not have such a multiplicative factor of $\o$. Now we impose the following commutation relations among the operators $a_\a$ and $b_\a$,
\begin{align}
\label{eq:accr}
\begin{split}
[a_\a, a_\b ] = & 0 = [b_\a, b_\b ], \\
[a_\a (\o, \z, \zb), b_\b (\o', \z', \zb')] &=  \d_{\a \b} \frac{\d(\o - \o')}{8\pi^2\o^3} \d(\z -\z', \zb - \zb').
\end{split}
\end{align}
Then using the integral representation of the step
function \bea \int_{-\infty}^{\infty} \frac{\exd \o}{\o} e^{\exi \o (u-u')} = 2
\pi \exi \Delta(u-u'), \eea we find that the commutation relations \eqref{eq:accr} ensure that 
the canonical commutation relations of $\cE$'s in Eq.~\eqref{eq:Eccr} are obeyed. \par
The relation between the operators $a$ and $b$ is obtained by imposing an appropriate
hermiticity condition. Since we are working in the stereographic coordinates $\{\z,\zb\}$, we require
that $(\cE_{\z})^\dagger = \cE_{\zb}$. Then using the reality of polarization vectors given in Eq.~\eqref{eq:polscri}
we get 
\bea
\label{eq:relab}
\begin{split}
b_+ = a^\dagger_- ,\\
b_-  = a^\dagger_+. 
\end{split}
\eea
Therefore, using Eqs.~(\ref{eq:Eccr}, \ref{eq:Efourier}, \ref{eq:relab}, \ref{eq:polscri}) we finally get the explicit form of the $\z$ and $\zb$ components of the electric-field 
operator as
\bea
\label{eq:Equantized}
\begin{split}
\cE_{\z} = \frac{\sqrt{2}}{(1+\z \zb)} \int_0 ^\infty \exd
\o \, \o\left[ a_{+}(\o, \z, \zb) \, e^{-\exi \o u} + a^\dagger_{-}(\o, \z, \zb) \, e^{\exi \o u}\right], \label{eq:Ezquant} \\
\cE_{\zb} = \frac{\sqrt{2}}{(1+\z \zb)} \int_0 ^\infty \exd
\o \, \o \left[ a_{-}(\o, \z, \zb) \, e^{-\exi \o u} + a^\dagger_{+}(\o, \z, \zb) \, e^{\exi \o u}\right]. \label{eq:Ezbquant}
\end{split}
\eea
From Eqs.~\eqref{eq:chargel} and \eqref{eq:Ezquant}, after doing the integral over $u$, we get the quantum
operator generating the large gauge symmetry with gauge parameter $\l$ as
\bea
\label{eq:chargeop}
Q_\l = \frac{-1}{2 \pi}  \int_{\mathbb{S}^2}  \exd^2\z \sqrt{q}  \,\, q^{\z \zb} \,  \exD_{\zb} \l \frac{\sqrt{2}}{(1+\z \zb)} \int_0 ^\infty \exd \o \, \o \d(\o) \left[  a_{+}(\o, \z, \zb) + a^\dagger_{-}(\o, \z, \zb)\right].
\eea
In particular, choosing the gauge parameter as $\l= \frac{1}{\z - z}$ and plugging in the metric \eqref{eq:stereometric},  either in Eq.~\eqref{eq:quantizableQ} or in 
Eq.~\eqref{eq:chargeop}, we get
\bea
\label{eq:softgen}
Q_{\l=\frac{1}{\z - z}} = -\frac{\sqrt{2}}{(1+z \bar{z})} \int_0 ^\infty \exd
\o\, \o \d(\o) \left[  a_{+}(\o, \z, \zb) + a^\dagger_{-}(\o, \z, \zb)\right].
\eea
 The explicit form of the gauge generator in Eq.~\eqref{eq:softgen} indicates that it creates/annihilates a photon with vanishing small energy. In the vacuum defined by $a_{\pm} |0\rangle = 0$, we see that the state obtained by the action of $Q_\l$ is not normalizable. Thus the large gauge symmetry is spontaneously broken in this vacuum. Soft photons are then the Goldstone modes of the spontaneously  broken large gauge symmetry \cite{Strominger:2013lka, He:2014cra}. 
\section{Promoting phase-space symmetry to the symmetry of S-matrix}
\label{sec:phtoS}
So far we have constructed the charge operator that generates the large gauge-symmetry on the radiative phase space constructed on $\fni$, i.e., it 
transforms one set of given initial data on $\fni$ to a new initial dataset. We could also have defined the initial data on the 
$\pni$ and there would be a corresponding generator of large gauge-symmetry on the corresponding phase space. Let us denote by $\cG^{+}$ ($\cG^{-}$) the group of large gauge symmetries at $\fni$ ($\pni$). Let the corresponding generators be $\cQ^{out}$ ($\cQ^{in}$). Each of these generators consists of a matter piece $Q_\l^{matter}$ and a radiation piece $Q_{\l}$,
\bea
\cQ^{out}_{\l} = Q_\l^{matter} + Q_{\l},
\eea
and similarly for $\cQ^{in}_\l$. The action of matter part on a state is given by
\bea
Q_\l^{matter}| \alpha \rangle = - \sum_{n \in \a}e_n \, \l(z_n, \bar{z}_n) \, | \alpha \rangle,
\eea
 where the sum involves the charge $e$ and the location $z$ (hence momentum) of each particle in the state $| \alpha \rangle$.\par
There is, in general, no relation between the gauge parameter on $\fni$ and that on $\pni$, we are free to choose them independently. Thus neither $\cG_+$ nor $\cG_-$ is the symmetry of the S-matrix. In order to promote the symmetry of the phase space to the symmetry of the S-matrix, Strominger suggested to first identify the null generators of $\fni$ and $\pni$ by the antipodal mapping of the $\mathbb{S}^2$ cross-sections and then equate the gauge parameters on $\fni$ and $\pni$ along the identified generators \cite{Strominger:2013jfa}. This can clearly be done in Minkowski space because an ingoing null ray originating at a point of  $\mathbb{S}^2$ of $\pni$ will hit the antipodal point of $\mathbb{S}^2$ at $\fni$. Presumably, this can also be done in spacetimes which are ``sufficiently close" to Minkowski \cite{Christodoulou:1993uv} (see also Ref.~\cite{Ludvigsen:92}). Identifying the generators of $\fni$ and $\pni$ in this way, and equating the gauge parameters on $\fni$ and $\pni$ we promote the diagonal subgroup $ \cG_{diag} \subset \cG_+ \times \cG_-$ of the large gauge-symmetry of phase space defined via the initial- and final-data surface to the symmetry of the S-matrix. Thus we obtain the following Ward identity \cite{Strominger:2013jfa}
\bea
\label{eq:Wardid}
\langle \b | \,\cQ^{out}_\l \,S - S \,\cQ^{in}_\l \,|\a \rangle = 0.
\eea
In the next section we will see that this Ward identity is equivalent to Weinberg's soft-photon theorem as written in Eq.~\eqref{eq:softni}. In what follows, we will suppress the tags $out/in$. It would be clear from the state on which the operator is acting whether it is the $out$ or $in$ type. 
\section{Ward identity $\iff$ Weinberg's soft-photon theorem}
\label{sec:wardsoft}
Separating the matter and radiation piece of $\cQ$ in Eq.~\eqref{eq:Wardid} we get,
\bea
\label{eq:Wardsep}
\langle \b | \,Q_\l \,S - S \,Q_\l \,|\a \rangle = - \langle \b | \,Q^{matter}_\l \,S - S \,Q^{matter}_\l \,|\a \rangle.
\eea
Using Eq.~\eqref{eq:chargeop} and the crossing symmetry $\langle \b |\,S\,a^\dagger_-(\o, \z, \zb)\,|\a \rangle = \langle \b |\,a_+(-\o,\z, \zb)\,S\,|\a \rangle$, the left-hand-side of Eq.~\eqref{eq:Wardsep} becomes,
\bea
\label{eq:wardlhs}
\langle \b | \,Q_\l \,S - S \,Q_\l \,|\a \rangle &=& \nn \\ 
\frac{-1}{2 \pi}  \int_{\mathbb{S}^2}  &\exd^2& \z \sqrt{q}  \,\, q^{\z \zb} \,  \exD_{\zb} \l \frac{\sqrt{2}}{(1+\z \zb)} \int_{-\infty} ^\infty \exd
\o \, \o \d(\o) \langle \b | \, a_{+}(\o, \z, \zb) \,S\,|\a \rangle. \nn \\
= \frac{-1}{2 \pi}  \int_{\mathbb{S}^2}  &\exd^2& \z \sqrt{q}  \,\, q^{\z \zb} \,  \exD_{\zb} \l \frac{\sqrt{2}}{(1+\z \zb)} \lim_{\o \to 0}  \langle \b | \, \o a_{+}(\o, \z, \zb) \,S\,|\a \rangle,
\eea
while the right-hand-side of Eq.~\eqref{eq:Wardsep} gives
\bea
\label{eq:wardrhs}
 - \langle \b | \,Q^{matter}_\l \,S - S \,Q^{matter}_\l \,|\a \rangle = \left[ \sum_{n \in \b }e_n \, \l(\z_n, \zb_n) - \sum_{n \in \a} e_n\,\l(\z_n, \zb_n)\,\right] \langle \b | \,S\,|\a \rangle.
\eea
Now plugging in the metric \eqref{eq:stereometric} and choosing the gauge parameter $\l (\z, \zb)= \frac{1}{\z - z}$ in Eqs.~(\ref{eq:wardlhs}, \ref{eq:wardrhs}), the Ward identity Eq.~\eqref{eq:Wardid} becomes,
\bea
\label{eq:almostsofttheorem}
\frac{\sqrt{2}}{(1+z \bar{z})} \lim_{\o \to 0}  \langle \b | \, \o a_{+}(\o, z, \bar{z}) \,S\,|\a \rangle = \left[ \sum_{n \in \b }e_n \, \frac{1}{z - \z_n} - \sum_{n \in \a} e_n\,\frac{1}{z - \z_n}\,\right] \langle \b | \,S\,|\a \rangle.
\eea
Rearranging, we get,
\bea
\lim_{\o \to 0}  \langle \b | \,a_{+}(\o, z, \bar{z}) \,S\,|\a \rangle = \lim_{\o \to 0} \frac{(1+z \bar{z})}{\sqrt{2}\o}\left[ \sum_{n \in \b }e_n \, \frac{1}{z - \z_n} - \sum_{n \in \a} e_n\,\frac{1}{z - \z_n}\,\right] \langle \b | \,S\,|\a \rangle,
\eea
and recognizing that $\lim_{\o \to 0}  \langle \b | \, a_{+}(\o, z, \bar{z}) \,S\,|\a \rangle = \lim_{\o \rightarrow 0} \mathscr{M}^{(+)}_{\alpha \rightarrow \beta}(q)$ we get the soft-photon theorem in the form written
in terms of quantities defined intrinsically on $\ni$ as in Eq.~\eqref{eq:softni}. This shows that Ward identity in Eq.~$\eqref{eq:Wardid}$ implies the soft-photon theorem in Eq.~\eqref{eq:Wsoft}. Finally, we  invoke the observation of Ref.~\cite{Strominger:2013lka} that any gauge parameter $\l(\z,\zb)$ can be written as 
\bea
\l(\z,\zb) = \frac{1}{2\pi} \int_{\mathbb{S}^2}  &\exd^2& z \sqrt{q}  \,\, q^{z \bar{z}} \, \l(z, \bar{z}) \exD_{\bar{z}} \frac{1}{z-\z}, \nn
\eea
to show that starting from Eq.~\eqref{eq:almostsofttheorem}, we can multilply both sides with a $\exD_{\zb} \,\l(\z, \zb)$ and retrace the steps back. Thus the Ward identity in Eq.~\eqref{eq:Wardid} for general large gauge parameter follows from the soft-photon theorem in Eq.~\eqref{eq:softni}. 
\section{Summary and Outlook}
\label{sec:summary}
In this paper we have derived the Weinberg's soft-photon theorem for massless Abelian gauge theory as a Ward identity corresponding to the diagonal subgroup of the group of large gauge symmetries acting on the radiative data  on $\fni$ and $\pni$. Our derivation is a slight generalization of that of Refs.~\cite{Strominger:2013lka, He:2014cra} because it also applies when the bulk spacetime is close to being Minkowskian, but is not exactly flat.  Our calculations are done on the conformal boundary $\ni$ of the spacetime. We have worked throughout with the quantities defined intrinsically on $\ni$. While this is in line with our motivation, which is holography for asymptotically flat space, it is far removed from the known examples of holography. For example, in AdS/CFT the boundary values of the bulk fields source the dual operators in the boundary gauge theory. One doesn't simply take the boundary values of the fields and quantize them. Nonetheless, we believe that our treatment of soft-theorem and Ward identity with a clean separation of the role of bulk and the boundary will be useful in developing holography in flat space. \par 
We have not discussed the subleading terms in the soft-photon theorem \cite{Low:1954kd, Low:1958sn}. It is known that the subleading soft-photon theorem implies a Ward identity \cite{Lysov:2014csa}. But what is the symmetry that this Ward identity corresponds to is not known. There could be a possibility that $\BMS$ and large-gauge transformations interact in such a way so as to generate more interesting symmetries that could account for the subleading terms in the Ward identity. But the algebra formed by $\BMS$ generators together with the generators of large-gauge symmetry discussed in Sec.~\ref{subsec:bmsaction}  
does not seem to have a rich structure to support this point of view. The  question thus remains open at the moment. The expression of $\BMS$ generators (also given in Ref.~\cite{Ashtekar:1981bq}) would nevertheless be useful in the study of the relationship between the soft-theorem and the electromagnetic memory effect as recently suggested in Ref.~\cite{Strominger:2014pwa}. The work along this direction is in progress and will be reported elsewhere. \par
%It is also worth mentioning that the although the symmetry group of massless QED in the bulk is the conformal group, the radiative phase space admits the action of the infinite dimensional $\BMS$ group.\par
We noted in Sec.~\ref{subsec:quantizeA} that when the large gauge-symmetry is admitted, the Poisson bracket of $A$'s is not well defined. We traced back the ``factor of 1/2" problem noted in Refs.~\cite{He:2014cra, He:2014laa} to the ambiguity in the Poisson bracket of $A$'s due to the presence of  large gauge-symmetry. Although one could avoid this problem by choosing electric field as the fundamental variable as we did in Sec.~\ref{subsec:quantizeE}, we considered it instructive to explain the resolution of this problem as provided in Refs.~\cite{He:2014cra, He:2014laa} via a toy model. The toy model makes it clear that one advantage of the approach in Refs.~\cite{He:2014cra, He:2014laa} is that it easily leads us to identify the soft-photons as edge states. In any case, one would need the operator corresponding to the vector potential in order to quantize, say, the $\BMS$ generators. Furthermore, the identification of soft-photons as edge states could potentially have important consequences.  We can not resist the temptation to speculate on the role of soft-modes in Hawking radiation and information loss in black-hole evaporation. The edge states carry entanglement entropy  \cite{Donnelly:2014fua, Huang:2014pfa, 
 Balachandran:1995iq}. Could it be that when one takes the entropy of the soft modes (say, of soft-gravitons) into account then the Hawking radiation is not thermal after all? We emphasize that this proposal is very different from the edge states associated to the presence of horizon (see, for e.g., Refs.~\cite{Carlip:1999cy, Balachandran:1994up}) where one associates the edge states to be localized on the horizon to account for the horizon entropy, i.e., horizon is the edge. Our suggestion, instead, is to calculate the contribution of soft modes to the entropy of Hawking radiation. The edge states in this case are the soft modes  and they lie on the boundary of $\fni$, i.e., on the ``edge of infinity" in some sense. \par
Finally, it should not be too difficult to  extend our analysis to non-Abelian gauge theories and gravity. 
%--application to firewall. Hilbert space decomposition is not early Hawking radiation + late Hawking radiation but early+late+edge states lying on the S^2 of scri. Too speculative!! well, should analyze the details to see if it works.
\begin{acknowledgments}
It is a pleasure to thank Tim Adamo, Joan Camps, Eduardo Casali, David Skinner and David Tong for stimulating discussions. We also thank Andrew Larkoski for comments and correspondence related to his work on soft-theorems. The final stages of this work were  completed while the author was visiting DAMTP as a Sciama Fellow through the generous support of the  Sciama Legacy Bursary.  The author is grateful to the members of DAMTP for their warm hospitality. 
\end{acknowledgments}

\appendix
\section{Symplectic formulation of field theory} \label{app:symplectic}
In this appendix we give a lightening review of the covariant phase space formulation of field theory in the the symplectic language. We refer the reader to Refs.~\cite{1987thyg.book..676C, bombelli1991covariant} for a detailed description. \par
The covariant phase space $\G$ is a symplectic manifold equipped with a closed
two-form $\O$ called the pre-symplectic structure. Each point of $\G$
is the solution of the equation of motion and thus represents the entire history of the system. Degenerate directions of $\O$ are
the gauge transformations of the theory. The degenerate directions can be shown
to be integrable. One can quotient $\G$ by these integral manifolds and obtain
the reduced phase space (also denoted by $\G$) which now inherits a
non-degenerate symplectic structure (also denoted by $\O$). \par The observables of the theory are represented by certain functions on $\G$. Every observable $f$ defines a flow on the phase space by specifying a
vector field $X_f$ associated to it as \bea \d f = \exi_{X_f} \O, \eea where
$\d$ denotes the exterior derivative on the phase space. Given two functions $f$
and $g$, the Poisson bracket between them is defined as \bea \{f,g\} :=
\O(X_f,X_g). \eea Poisson bracket satisfies the Jacobi identity, i.e, for three
observables, $f,g,h$ we have \bea \{f ,\{g , h\}\} + \{g ,\{ h,f \}\} + \{h ,\{f
,g \}\} = 0. \eea The vector field $X_f$ generated by a function $f$ acts on a
function $g$ as \bea X_f(g) := \{f,g\}. \eea Hence, by definition, we have $X_f(g)
= -X_g(f)$. The Lie bracket between the vector fields can be calculated using
the Jacobi identity and we get \bea [X_f,X_g ] = X_{\{f,g\}}. \eea Those vector
fields on the phase space which Lie- drag $\O$ are called the symmetries of the theory. That is,
a vector field $X_h$ on phase space is said to generate a symmetry if \bea
\Lie_{X_h} \O = 0. \eea Since, $\O$ is closed we have that $\Lie_{X_h} \O = \d
\exi_{X_h} \O$, where we have used the identity $\Lie_{X_h} = \d \exi_{X_h} + \exi_{X_h} \d$. A necessary and sufficient condition for this to vanish is that
$\exi_{X_h} \O$ be an exact form, i.e., there is a function $h$ on phase space
such that \bea \d h = \exi_{X_h} \O. \eea Thus, the symmetries are associated with certain functions on the phase space. These functions thus associate a conserved quantity to each history, for e.g., the energy associated to a spacetime is the value of the Hamiltonian function. 

\subsection*{An example: Free Maxwell field} Let us now see how it all works out in the free
Maxwell field. The action is given by \bea S = \frac{-1}{4} \int \exd^4 x
\sqrt{-g} \, F_{\mu \nu} F^{\mu \nu}, \eea where $F_{\mu \nu}$ is the field
strength of the U(1) gauge field $A_\mu$. Variation of the action is given by
\bea \d S= \int \exd^4 x \sqrt{-g} \, \d A_\nu \, \n_\mu F^{\mu \nu} - \oint
\exd^3 x \sqrt{h} \,n_\mu F^{\mu \nu} \, \d A_\nu, \eea where $h$ is the
determinant of the induced metric $h_{\mu \nu}$ on the boundary and $n_\mu$ is
its covariant normal. Bulk term yields the equation of motion $\n_\mu F^{\mu
\nu} = 0$. Integral of the boundary integrand on a constant time-slice $\Sigma$
defines the symplectic potential, \bea \Theta = - \int_\Sigma \exd^3 x \,
\sqrt{h}\, n_\mu F^{\mu \nu}\, \d A_\nu. \eea Pre-symplectic structure is now
given by the exterior derivative of the symplectic potential as \bea \O &=& \d
\Theta, \\ &=& - \int_\Sigma \exd^3 x \sqrt{h}\, \d E^\mu \, \d A_\mu, \eea
where $\d E^\mu := n_\mu \d F^{\mu \nu}$. We thus see that the phase space is
coordinatized by $\{A_\mu,E^\mu\}$. Now, consider the vector field on the phase
space given by \bea X_\Lambda = \int \exd^4 x \, \n_\mu\Lambda \, \frac{\d}{\d
A_\mu} \eea where $\Lambda$ is a spacetime scalar. If $\exi_X \O= 0$ then this
is the gauge direction of $\O$. Calculating \bea \exi_{X_\Lambda} \O = \int
\exd^3 x \,\sqrt{h} \, \d E^\mu \n_\mu \Lambda. \eea Now from the definition of
the Electric field we see that $n_\mu E^\mu = 0$, hence the derivative of
$\Lambda$ is really only in the direction orthogonal to $n^\mu$. Denoting the
intrinsic covariant derivative on the slice $\Sigma$ as $D_\mu$, we get after
integration by parts \bea \exi_{X_\Lambda} \O = - \int \exd^3 x \sqrt{h} (D_\mu
\,\d E^\mu) \, \Lambda + \oint \exd^2 x \,\sqrt{q} \, m_\mu \d E^\mu \Lambda,
\eea where $m_\mu$ is the covariant normal to the boundary of $\Sigma$ at
infinity. Now, projecting the equation of motion $\n_\mu F^{\mu \nu}$ on the
slice $\Sigma$ we get the Gauss law $D_\mu E^\mu = 0$, which is interpreted as a
constraint in the canonical theory. In the covariant phase space of course there
are no constraints. Using the linearized equation $D_\mu \d E^\mu = 0$, we get
\bea \exi_{X_\Lambda} \O = \oint \exd^2 x \,\sqrt{q} \, m_\mu \d E^\mu \Lambda.
\eea Thus we see that only when $\Lambda$ vanishes on the boundary of $\Sigma$
that $X_\Lambda$ is a degenerate direction of $\O$ and the motion along $X_\Lambda$ is
to be interpreted as gauge. If $\Lambda$ does not vanish on the boundary of
$\Sigma$ then we define the hamiltonian or charge $Q_\Lambda$ as $\d Q_\Lambda =
\exi_{X_\Lambda} \O$ where $Q_\Lambda$ is given by \bea Q_\Lambda = \int \exd^2
x \sqrt{q} \,m_\mu E^\mu \,\Lambda. \eea The charge $Q_\Lambda$ generates the transformation $A_\mu \to A_\mu + \nabla_\mu \Lambda$ on the phase space. For the algebra of charges,  we  find that
 $\{ Q_{\Lambda_1},Q_{\Lambda_2} \} = 0$, hence the charge algebra is abelian. 

\bibliography{softYMref}

\end{document}
%
% ****** End of file apstemplate.tex ******